\documentclass{eptcs-modified}
 \providecommand{\event}{} 
\usepackage{amsmath}
\usepackage{amssymb}
\usepackage{amsthm}
\usepackage{amscd}
\usepackage{graphics}
\usepackage{soul}

\long\def\greybox#1{%
    \newbox\contentbox%
    \newbox\bkgdbox%
    \setbox\contentbox\hbox to \hsize{%
        \vtop{
            \kern\columnsep
            \hbox to \hsize{%
                \kern\columnsep%
                \advance\hsize by -2\columnsep%
                \setlength{\textwidth}{\hsize}%
                \vbox{
                    \parskip=\baselineskip
                    \parindent=0bp
                    #1
                }%
                \kern\columnsep%
            }%
            \kern\columnsep%
        }%
    }%
    \setbox\bkgdbox\vbox{
        \pdfliteral{0.85 0.85 0.85 rg}
        \hrule width  \wd\contentbox %
               height \ht\contentbox %
               depth  \dp\contentbox
        \pdfliteral{0 0 0 rg}
    }%
    \wd\bkgdbox=0bp%
    \vbox{\hbox to \hsize{\box\bkgdbox\box\contentbox}}%
    \vskip\baselineskip%
}

\title{Effective local compactness and the hyperspace of located sets}
\author{
Arno Pauly
\institute{Swansea University\\Swansea, UK}
\email{Arno.M.Pauly@gmail.com}
}
\def\titlerunning{Effective local compactness}
\def\authorrunning{A. Pauly}

\begin{document}
\theoremstyle{definition}
\newtheorem{theorem}{Theorem}
\newtheorem{definition}[theorem]{Definition}
\newtheorem{problem}[theorem]{Problem}
\newtheorem{assumption}[theorem]{Assumption}
\newtheorem{corollary}[theorem]{Corollary}
\newtheorem{proposition}[theorem]{Proposition}
\newtheorem{lemma}[theorem]{Lemma}
\newtheorem{observation}[theorem]{Observation}
\newtheorem{fact}[theorem]{Fact}
\newtheorem{question}[theorem]{Open Question}
\newtheorem{conjecture}[theorem]{Conjecture}
\newtheorem{example}[theorem]{Example}
\newtheorem{remark}[theorem]{Remark}
\newcommand{\dom}{\operatorname{dom}}
\newcommand{\id}{\textnormal{id}}
\newcommand{\Cantor}{{\{0, 1\}^\mathbb{N}}}
\newcommand{\Baire}{{\mathbb{N}^\mathbb{N}}}
\newcommand{\Lev}{\textnormal{Lev}}
\newcommand{\hide}[1]{}
\newcommand{\mto}{\rightrightarrows}
\newcommand{\uint}{{[0, 1]}}
\newcommand{\bft}{\mathrm{BFT}}
\newcommand{\lbft}{\textnormal{Linear-}\mathrm{BFT}}
\newcommand{\pbft}{\textnormal{Poly-}\mathrm{BFT}}
\newcommand{\sbft}{\textnormal{Smooth-}\mathrm{BFT}}
\newcommand{\ivt}{\mathrm{IVT}}
\newcommand{\cc}{\textrm{CC}}
\newcommand{\lpo}{\textrm{LPO}}
\newcommand{\llpo}{\textrm{LLPO}}
\newcommand{\aou}{AoU}
\newcommand{\Ctwo}{C_{\{0, 1\}}}
\newcommand{\name}[1]{\textsc{#1}}
\newcommand{\C}{\textrm{C}}
\newcommand{\CC}{\textrm{CC}}
\newcommand{\UC}{\textrm{UC}}
\newcommand{\ic}[1]{\textrm{C}_{\sharp #1}}
\newcommand{\xc}[1]{\textrm{XC}_{#1}}
\newcommand{\me}{\name{P}.~}
\newcommand{\etal}{et al.~}
\newcommand{\eval}{\operatorname{eval}}
\newcommand{\rank}{\operatorname{rank}}
\newcommand{\Sierp}{Sierpi\'nski }
\newcommand{\isempty}{\operatorname{IsEmpty}}
\newcommand{\spec}{\textrm{Spec}}
\newcommand{\cord}{\textrm{COrd}}
\newcommand{\Cord}{\textrm{\bf COrd}}
\newcommand{\CordM}{\Cord_{\textrm{M}}}
\newcommand{\CordK}{\Cord_{\textrm{K}}}
\newcommand{\CordHL}{\Cord_{\textrm{HL}}}
\newcommand{\leqW}{\leq_{\textrm{W}}}
\newcommand{\leW}{<_{\textrm{W}}}
\newcommand{\equivW}{\equiv_{\textrm{W}}}
\newcommand{\geqW}{\geq_{\textrm{W}}}
\newcommand{\pipeW}{|_{\textrm{W}}}
\newcommand{\nleqW}{\nleq_{\textrm{W}}}
\newcommand{\Det}{\textrm{Det}}
\newcommand{\tktm}{$\mathrm{T}2\kappa\mathrm{TM}$}

\newcommand\tboldsymbol[1]{%
\protect\raisebox{0pt}[0pt][0pt]{%
$\underset{\widetilde{}}{\boldsymbol{#1}}$}\mbox{\hskip 1pt}}

\newcommand{\bolds}{\tboldsymbol{\Sigma}}
\newcommand{\boldp}{\tboldsymbol{\Pi}}
\newcommand{\boldd}{\tboldsymbol{\Delta}}
\newcommand{\boldg}{\tboldsymbol{\Gamma}}

\newcounter{saveenumi}
\newcommand{\seti}{\setcounter{saveenumi}{\value{enumi}}}
\newcommand{\conti}{\setcounter{enumi}{\value{saveenumi}}}

\maketitle

\begin{abstract}

\end{abstract}


\section{Introduction}
We revisit the question of how to effectivize the notion of local compactness. Our approach is in the traditions of both synthetic topology \cite{escardo} (taking the notion of continuous function as a primitive, and employing category-theoretic machinery) and computable topology. Computable topology explores the effective counterparts to definitions and theorems of general topology, which serve in particular as part of the foundations of computable analysis.

There have been previous studies of effective local compactness, albeit restricted to computable Polish spaces. We compare the definitions in Subsection \ref{subsec:comparison}, and show that they are equivalent. There are some subtleties involved, which could be interpreted as demonstrating that the previous definitions were (even for computable Polish spaces) prima facie too restrictive (as for some examples establishing their effective local compactness would be more work than it should be).

While our definition of effective local compactness (Definition \ref{def:main}) works for arbitrary represented spaces, to make good use the notion we quickly find ourselves desiring additional structure. We thus restrict our attention to countably-based spaces from Section \ref{sec:countablybased} onwards, and obtain the notion of an effectively relatively compact system (ercs) in Definition \ref{def:ercs} as the combinatorial underpinning of effective local compactness in countably-based spaces. In, in addition, we demand our spaces to be (computably) Hausdorff, we enter the realm of (computably) metrizable spaces, which we explore in Section \ref{sec:metric}.

As application of effective local compactness, in Section \ref{sec:av} we study the space $(\mathcal{A} \wedge \mathcal{V})(\mathbf{X})$ of closed-and-overt subsets of a given space (or of \emph{located sets} or of \emph{closed subsets with full information}, depending on the nomenclature). We show that $\mathbf{X}$ admitting an ercs suffices to make $(\mathcal{A} \wedge \mathcal{V})(\mathbf{X})$ computably compact (Corollary \ref{corr:ercscompact}) and computably metrizable (Corollary \ref{corr:metrizable}). This generalizes a previous result: \cite{ziegler9} shows that for a computably compact computable metric space $\mathbf{X}$, the space $(\mathcal{A}\wedge\mathcal{V})(\mathbf{X}) \cong (\mathcal{K}\wedge\mathcal{V})(\mathbf{X})$ is a computably compact computable metric space again. In the restricted setting, the metric on $(\mathcal{A} \wedge \mathcal{V})(\mathbf{X})$ is the Hausdorff distance. Since \cite{ziegler9} thus can work with a known construction of the metric, whereas we employ Schr\"oder's effective metrization theorem, the proofs are very different.

To establish $(\mathcal{A} \wedge \mathcal{V})(\mathbf{X})$ as computably compact means that universal quantification over closed-and-overt sets preserves open predicates, following the characterization of computable compactness in \cite{pauly-synthetic}. This gives a clear indication of how effective local compactness can be used.

\section{Defining effective local compactness}
For non-Hausdorff spaces, there are several competing (and non-equivalent) definitions of local compactness, see e.g.~the overview provided in the Wikipedia article \cite{wiki-locallycompact}. We will effective the existence of a compact local neighborhood basis for our purposes:

\begin{definition}
\label{def:main}
We call a represented space $\mathbf{X}$ \emph{effectively locally compact}, if the map
$$\mathrm{CompactBase} : \subseteq \mathbf{X} \times \mathcal{O}(\mathbf{X}) \mto \mathcal{O}(\mathbf{X}) \times \mathcal{K}(\mathbf{X})$$
with $\dom(\mathrm{CompactBase}) = \{(x,U) \mid x \in U\}$ and $(V,K) \in \mathrm{CompactBase}(x,U)$ iff $x \in V \subseteq K \subseteq U$ is computable.
\end{definition}

Just as for the classical notion, we see that effective local compactness is preserved by taking open or closed subspaces:

\begin{proposition}
\label{prop:closedsubspace}
Let $A \in \mathcal{A}(\mathbf{X})$ be computable and $\mathbf{X}$ be effectively locally compact. Then the subspace $\mathbf{A}$ of $\mathbf{X}$ is effectively locally compact.
\begin{proof}
We are given $x \in \mathbf{A}$ and $U \in \mathcal{O}(\mathbf{A})$. We can compute $U \cup A^C \in \mathcal{O}(\mathbf{X})$. We apply $\mathrm{CompactBase}_\mathbf{X}(x,U \cup A^C)$ to obtain some $(V,K) \in \mathcal{O}(\mathbf{X}) \times \mathcal{K}(\mathbf{X})$. From that, we can compute $V \cap U \in \mathcal{O}(\mathbf{A})$, and $K \cap A \in \mathcal{K}(\mathbf{A})$, and these are a valid output to $\mathrm{CompactBase}_\mathbf{A}(x,U)$.
\end{proof}
\end{proposition}

\begin{proposition}
\label{prop:opensubspace}
Let $Y \in \mathcal{O}(\mathbf{X})$ be computable and $\mathbf{X}$ be effectively locally compact. Then the subspace $\mathbf{Y}$ of $\mathbf{X}$ is effectively locally compact.
\begin{proof}
For the computably open subspace $\mathbf{Y}$ we have the canonic computable embedding $\id : \mathcal{O}(\mathbf{Y}) \to \mathcal{O}(\mathbf{X})$. From that, we also get that $\id : \subseteq \mathcal{K}(\mathbf{X}) \to \mathcal{K}(\mathbf{Y})$ is computable. Together, these yield the claim.
\end{proof}
\end{proposition}

\begin{corollary}
\label{corr:subspace}
Let $\mathbf{X}$ be effectively locally compact, $A \in \mathcal{A}(\mathbf{X})$ be computable and $Y \in \mathcal{O}(\mathbf{X})$ be computable. Then the subspace $\mathbf{A} \cap \mathbf{Y}$ of $\mathbf{X}$ is effectively locally compact.
\begin{proof}
Note that $A \cap Y$ is a computably open subset of $\mathbf{A}$, and combine Propositions \ref{prop:closedsubspace}, \ref{prop:opensubspace}.
\end{proof}
\end{corollary}

Taking more general subspaces does not preserve local compactness. In fact, any locally compact subspace of a Hausdorff space is the intersection of an open and a closed subspace (see e.g.~\cite{mathse-locallycompact}). As such, Corollary \ref{corr:subspace} already realizes the full extent of what we can hope for.

\hide{
PROBLEM HERE WITH TAKING INTERSECTION INDEXED BY PARTIAL MULTIVALUED FUNCTIONS

Classically, it is trivial to see that locally compact spaces are closed under finite products (and they are not closed under infinite products). However, the argument invokes that the product topology is generated by sets of the form $U \times V$ where $U$ and $V$ are open. This does not hold for the product of represented spaces, however, since the topology on $\mathbf{X} \times \mathbf{Y}$ is the sequentialization of the product topology instead. A slightly more involved argument still yields the desired result though:

\begin{theorem}
Let $\mathbf{X}$ and $\mathbf{Y}$ be effectively locally compact. Then so is $\mathbf{X} \times \mathbf{Y}$.
\begin{proof}
We are given $(x,y) \in \mathbf{X} \times \mathbf{Y}$ and $U \in \mathcal{O}(\mathbf{X} \times \mathbf{Y})$ with $x \in U$. First, we compute $\mathrm{Cut}_\mathbf{X}(x,U) = \{y' \in \mathbf{Y} \mid (x,y') \in U\} \in \mathcal{O}(\mathbf{Y})$, and note that $(x,\mathrm{Cut}_\mathbf{X}(x,U))$ is a valid input for $\mathrm{CompactBase}_\mathbf{Y}$. This yields some $V \in \mathcal{O}(\mathbf{Y})$ and $A \in \mathcal{K}(\mathbf{Y})$ with: $$(x,y) \in \{x\} \times V \subseteq \{x\} \times A \subseteq U$$
We now proceed relative to some parameter $z \in A$, and compute $\mathrm{Cut}_\mathbf{Y}(z,U) \in \mathcal{O}(\mathbf{X})$, and note that we can apply $\mathrm{CompactBase}_\mathbf{X}$ to $(z,\mathrm{Cut}_\mathbf{Y}(z,U))$ to obtain $W_z \in \mathcal{O}(\mathbf{X})$ and $B_z \in \mathcal{K}(\mathbf{X})$ with: $$(x,z) \in W_z \times \{z\} \subseteq B_z \times \{x\} \subseteq U$$
From this, we can conclude that also:
$$(x,y) \in \left (\bigcap_{z \in A} W_z \right ) \times V \subseteq \left (\bigcap_{z \in A} B_z \right ) \times A \subseteq U$$
Since the dependence of $W_z$ and $B_z$ on the parameter $z$ is continuous, and both open and compact sets are effectively closed under compact intersections, we see that we can actually compute $\left (\bigcap_{z \in A} W_z \right ) \in \mathcal{O}(\mathbf{X})$ and $\left (\bigcap_{z \in A} B_z \right ) \in \mathcal{K}(\mathbf{X})$. With that, we can compute $\left (\bigcap_{z \in A} W_z \right ) \times V \in \mathcal{O}(\mathbf{X} \times \mathbf{Y})$ and $\left (\bigcap_{z \in A} B_z \right ) \times A \in \mathcal{K}(\mathbf{X} \times \mathbf{Y})$, and these constitute a valid output to $\mathrm{CompactBase}_{\mathbf{X} \times \mathbf{Y}}((x,y),U)$.
\end{proof}
\end{theorem}
}

\section{Effective local compactness for countably-based spaces}
\label{sec:countablybased}

In countably-based spaces, we can ask for a specific structure that witnesses effective local compactness. Manipulating this structure will be how we prove further results.
\begin{definition}
\label{def:ercs}
Let an effective relatively compact system (ercs) of a represented space be a triple $((U_n)_{n \in \mathbb{N}},(B_n)_{n \in \mathbb{N}},R)$ where
\begin{enumerate}
\item $(U_n \in \mathcal{O}(\mathbf{X}))_{n \in \mathbb{N}}$ is a computable sequence of open sets;
\item $(B_n \in \mathcal{K}(\mathbb{N}))_{n \in \mathbb{N}}$ is a computable sequence of compact sets;
\item and $R \subseteq \mathbb{N} \times \mathbb{N}$ is a computably enumerable relation such that $(m,n) \in R$ implies $U_m \subseteq B_n$;
\end{enumerate}
such that for any open set $U \in \mathcal{O}(\mathbf{X})$ it holds that:
$$U = \bigcup_{\{n \mid U \supseteq B_n\}} \bigcup_{\{m \mid (m,n) \in R\}} U_m$$
\end{definition}

The idea is that $R$ codes a formal containment relation between the enumerated open and compact sets. We shall write $U_n \ll B_m$ for $(n,m) \in R$.

\begin{proposition}
\label{prop:countablebasis}
Let $((U_n)_{n \in \mathbb{N}},(B_n)_{n \in \mathbb{N}},R)$ be an ercs of $\mathbf{X}$. Then $(U_n)_{n \in \mathbb{N}}$ is an effective countable basis of $\mathbf{X}$.
\begin{proof}
We are given some $x \in \mathbf{X}$, $U \in \mathcal{O}(\mathbf{X})$ with $x \in U$, and we are searching for some $n \in \mathbb{N}$ such that $x \in U_n \subseteq U$. By assumption, we have that:
$$U = \bigcup_{\{k \mid U \supseteq B_k\}} \bigcup_{\{n \mid (n,k) \in R\}} U_n$$
Thus, $x \in U$ implies that there are $k,n$ with $B_k \subseteq U$, $U_n \ll B_k$ and $x \in U_n$. By definition of compact sets, we can effectively enumerate all $k$ such that $B_k \subseteq U$, by definition of $\ll$ we can enumerate all $n$ such that $U_n \ll B_k$, and by definition of open sets we can enumerate all $n$ such that $x \in U_n$. We will thus eventually find $k,n$ as above, and can then output $n$.
\end{proof}
\end{proposition}

Having an ercs is a form of effective local compactness, as witnessed by:

\begin{proposition}
\label{prop:ercsimplieslc}
Let $\mathbf{X}$ have an ercs. Then the $\mathbf{X}$ is effectively locally compact, i.e.~the map
$$\mathrm{CompactBase} : \subseteq \mathbf{X} \times \mathcal{O}(\mathbf{X}) \mto \mathcal{O}(\mathbf{X}) \times \mathcal{K}(\mathbf{X})$$
with $\dom(\mathrm{CompactBase}) = \{(x,U) \mid x \in U\}$ and $(V,K) \in \mathrm{CompactBase}(x,U)$ iff $x \in V \subseteq K \subseteq U$ is computable.
\begin{proof}
We proceed as in Proposition \ref{prop:countablebasis}, but output both $U_n$ as $V$ and $B_k$ as $K$.
\end{proof}
\end{proposition}

Conversely, the existence of en effective countable basis as in Proposition \ref{prop:countablebasis} and effective local compactness together are almost enough to imply the existence of an ercs -- we only need a mild additional constraint:

\begin{proposition}
\label{prop:converse}
Let $\mathbf{X}$ admit an effective countable basis $(U_n)_{n \in \mathbb{N}}$, a representation $\delta$ with a computable dense sequence $(p_n)_{n \in \mathbb{N}}$ in $\dom(\delta)$ (\footnote{To be precise, we only need that $\dom(\delta)$ is computably overt, and that there is a computable dense sequence $(x_n)_{n \in \mathbb{N}}$ in $\mathbf{X}$ for our proof. It is clear that we can construct specific representations fulfilling the more specific criteria, but not the general ones. It is less clear whether this can be extended to an entire equivalence class.}) and have the map
$$\mathrm{CompactBase} : \subseteq \mathbf{X} \times \mathcal{O}(\mathbf{X}) \mto \mathcal{O}(\mathbf{X}) \times \mathcal{K}(\mathbf{X})$$
be computable. Then we can construct $(B_n \in \mathcal{K}(\mathbb{N}))_{n \in \mathbb{N}}$ and $R \subseteq \mathbb{N} \times \mathbb{N}$ to make an ercs.
\begin{proof}
First, we construct the $B_k$. For that, set $x_\ell := \delta(p_\ell)$. We go through all $x_\ell$, $U_n$ with $x_\ell \in U_n$ and call $\mathrm{CompactBase}(x_\ell,U_n)$ to obtain some $V_{\ell n} \in \mathcal{O}(\mathbf{X})$ and $B_{\ell n} \in \mathcal{K}(\mathbf{X})$ with $x_\ell \in V_{\ell n} \subseteq B_{\ell n} \subseteq U_n$. Since $(U_n)_{n \in \mathbb{N}}$ is an effective countable basis, we can find some $m$ such that $x_\ell \in U_m \subseteq V_{\ell n} \subseteq B_{\ell n} \subseteq U_n$. For any combination of $n, m$ arising in this way, we pick some $k(n,m)$ and set $B_{k} = B_{\ell n}$.

Now, we construct $R$. Given some $p \in \dom(\delta)$ and some $n \in \mathbb{N}$ with $\delta(p) \in U_n$, we can apply first $\mathrm{CompactBase}(\delta(p),U_n)$ to obtain compact $K$ and open $V$ with $\delta(p) \in V \subseteq K \subseteq U_n$, and then use the fact that $(U_n)_{n \in \mathbb{N}}$ is an effective countable basis to find some $m \in \mathbb{N}$ with $\delta(p) \in U_m \subseteq V$. We only consider the map $(p,n) \mapsto m$. This has a single-valued computable choice function $\chi$. We can thus construct the open sets $O_{nm} = \{p \in \dom(\delta) \mid \delta(p) \in U_n \wedge \chi(p,n) = m\}$ of $\dom(\delta)$. Since $\dom(\delta)$ contains a computable dense sequence, it is computably overt. Thus, defining $(m,k(n,m)) \in R \Leftrightarrow O_{nm} \neq \emptyset$ yields a computably enumerable relation, which by construction satisfies that $U_i \ll B_j \Rightarrow U_i \subseteq B_j$.

It remains to show the main property of an ercs. It suffices to do so for basic open sets $U_n$. Moreover, the right-to-left inclusion is trivial. Thus, we start with some $x \in U_n$. Then there is some name $p \in \dom(\delta)$ with $\delta(p) = x$. Note that by construction, we have that $\bigcup_{m \in \mathbb{N}} O_{nm} = \delta^{-1}(U_n)$, thus there exists some $m$ such that $p \in O_{nm}$. But then $x \in U_m \subseteq B_{k(n,m)} \subseteq U_n$, and $U_m \ll B_{k(n,m)}$, and thus $x$ is present on the right hand side of the main property.
\end{proof}
\end{proposition}

\begin{remark}
By dropping any computability requirements from Proposition \ref{prop:converse}, we can conclude that any countably-based locally compact space will admit an ercs relative to some oracle. In particular, the notion of admitting an ercs passes the fundamental sanity check for being a notion of effective local compactness for countably based spaces.
\end{remark}

We briefly explore how admitting an ercs, being compact, and being computably compact are related:

\begin{proposition}
\label{prop:compactcomputablycompact}
Let $\mathbf{X}$ admit an ercs and be compact. Then $\mathbf{X}$ is computably compact.
\begin{proof}
By the main property of an ercs, we find that $\mathbf{X} = \bigcup_{k \in \mathbb{N}} \bigcup_{\{\ell \mid (\ell,k) \in R\}} U_\ell$. This is an open cover of $\mathbf{X}$. Compactness of $\mathbf{X}$ implies that there is a finite subcover, and that in particular there is a finite set $K \subseteq \mathbb{N}$ with $\mathbf{X} = \bigcup_{k \in K} \bigcup_{\{\ell \mid (\ell,k) \in R\}} U_\ell$. But this implies $\mathbf{X} = \bigcup_{k \in K} B_k$, and computably compact sets are closed under finite union.
\end{proof}
\end{proposition}

\begin{example}
There is an effectively countably-based computably compact space without an ercs (which is, in fact, not locally compact at all).
\begin{proof}
Let $\hat{\mathbb{Q}}$ be a one-point compactification of $\mathbb{Q}$ (with the Euclidean topology). This means that the underlying set is $\mathbb{Q} \cup \{\infty\}$, and the topology is generated by open subsets of $\mathbb{Q}$ together with sets of the form $\{\infty\} \cup (\mathbb{Q} \setminus I)$ for finite sets $I$.

In terms of representation, a name of a point $x \in \hat{\mathbb{Q}}$ starts with listing some rationals $q_0$, $q_1$, $q_2$, guaranteeing that $x \neq q_i$. Either this enumeration continues for ever and exhausts all rationals, in which case $x = \infty$. Alternatively, we reach a special stop flag, and after that read a usual $\mathbb{Q}$-name of a point different from the rationals enumerated so far.

This space is easily seen to be computably compact (after all, covering $\infty$ leaves only finitely many points to check). However, the open set $\mathbb{Q} \subseteq \hat{\mathbb{Q}}$ contains only compact sets with empty interior, there cannot be an ercs.
\end{proof}
\end{example}

\hide{
\subsection{$\mathcal{O}(\mathbf{X})$ as retract of $\mathcal{O}(\mathbb{N})$}
Recall that a space $\mathbf{X}$ is a computable retract of a space $\mathbf{Y}$, if there are computable function $f : \mathbf{X} \to \mathbf{Y}$ and $g : \mathbf{Y} \to \mathbf{X}$ with $\id_\mathbf{X} = g \circ f$.

\begin{theorem}
The following are equivalent for a sober represented space $\mathbf{X}$:
\begin{enumerate}
\item $\mathcal{O}(\mathbf{X})$ is a computable retract of $\mathcal{O}(\mathbb{N})$.
\item $\mathbf{X}$ admits an ercs.
\end{enumerate}
\begin{proof}
If $\mathbf{X}$ admits an ercs, then the computable maps $U \mapsto \{m \in \mathbb{N} \mid \exists k \in \mathbb{N} \ U_m \ll B_k \subseteq U\} : \mathcal{O}(\mathbf{X}) \to \mathcal{O}(\mathbb{N})$ and $V \mapsto \bigcup_{n \in V} U_n : \mathcal{O}(\mathbb{N}) \to \mathcal{O}(\mathbf{X})$ witness that $\mathcal{O}(\mathbf{X})$ is a computable retract of $\mathcal{O}(\mathbb{N})$. Note that the main condition of an ercs states precisely that the composition of the two maps is $\id_{\mathcal{O}(\mathbf{X})}$.

Now assume conversely that $\mathcal{O}(\mathbf{X})$ is a computable retract of $\mathcal{O}(\mathbb{N})$ witnessed by $f : \mathcal{O}(\mathbf{X}) \to \mathcal{O}(\mathbb{N})$ and $g : \mathcal{O}(\mathbb{N}) \to \mathcal{O}(\mathbf{X})$.

Pick some effective enumeration $(I_n)_{n \in \mathbb{N}}$ of the finite subsets of $\mathbb{N}$. We say that $I_n$ entails $I_m$ ($I_n \sqsupseteq I_m$), if there are $I_k, I_\ell$ with $I_k \cap I_\ell \subseteq I_n$, $f(g(I_k)) \supseteq I_m \subseteq f(g(I_\ell))$. Let $U_n = g(I_n)$ and $B_k = \{U \in \mathcal{O}(\mathbf{X}) \mid \exists I_n \subseteq f(U), I_n \sqsupseteq I_k\}$.

Claim: $B_k$ defines a compact set. We show that $B_k$ is a filter, and invoke the Hofman-Mislove theorem \cite{mislove}, which states that these define the compact sets. Being upwards-closed follows from the definition, so we only need to check being intersection-closed. Assume that $U, V \in \mathcal{O}(\mathbf{X})$ are such that $\exists I_n \subseteq f(U)$, $\exists I_\ell \subseteq f(V)$ with $I_n \sqsupseteq I_k$ and $I_\ell \sqsupseteq I_k$. TO BE COMPLETED

We define $\ll$ by $U_n \ll B_k$ iff $n = k$. The main property of an ercs now is the conclusion of Lemma \ref{lemma:retractpointwise} below.
\end{proof}
\end{theorem}

\begin{lemma}
\label{lemma:retractpointwise}
Let $f : \mathcal{O}(\mathbf{X}) \to \mathcal{O}(\mathbb{N})$ and computable $g : \mathcal{O}(\mathbb{N}) \to \mathcal{O}(\mathbf{X})$ satisfy $\id_{\mathcal{O}(\mathbf{X})} = g \circ f$. Then for all $U \in \mathcal{O}(\mathbf{X})$ it holds that $U = \bigcup_{n \in f(U)} g(\{n\})$.
\begin{proof}
As a computable function, $g$ is in particular Scott-continuous, i.e.~satisfies that $\bigcup_{n \in f(U)} g(\{n\}) = g(f(U))$.
\end{proof}
\end{lemma}
}

\section{Effective local compactness in computable metric spaces}
\label{sec:metric}

If a space admits an ercs and is computably Hausdorff, it is already computably metrizable. This follows very directly from Schr\"oder's effective metrization theorem \cite{schroder8,grubba3}. The latter states that computably regular effectively countably-based spaces are computably metrizable. Their formulation of being computably regular actually takes the very same form as the definition of ercs, except that closed sets are used in the place of compact sets. Since being computably Hausdorff suffices to translate from compact sets to closed sets, it follows that a computably Hausdorff space admitting an ercs is already computably regular.

Being computably metrizable is strictly more general than being a computable metric space, although every computably metrizable space embeds into a computable metric space. Still, this shows that restricting to computable metric space is not that restrictive in the context of effective local compactness. We will see that we can say a few more things using the language of metric spaces.

\subsection{Finding compact balls}
In computable metric spaces, we can be more specific regarding how the sets $B_n$ in ercs look like; namely, we can demand that the compact sets be closed\footnote{We need to carefully distinguish between the closed balls $\overline{B}(x,r)$ and the closures of open balls $\mathrm{cl} B(x,r)$ in this context. In general, the former are available as members of $\mathcal{A}(\mathbf{X})$, or, in the compact case, $\mathcal{K}(\mathbf{X})$. The latter are available as members of $\mathcal{V}(\mathbf{X})$.} balls:

\begin{proposition}
\label{prop:inmetricspaces}
Let $(\mathbf{X},d)$ be a computable metric space. Then the following are equivalent:
\begin{enumerate}
\item $\mathbf{X}$ admits an ercs.
\item The map $\operatorname{CompactBall} : \mathbf{X} \mto (\mathbb{N} \times \mathcal{K}(\mathbf{X}))$ where $(n,K) \in \operatorname{CompactBall}(x)$ iff $K = \overline{B}(x,2^{-n})$ is well-defined and computable.
\end{enumerate}
\begin{proof}
\begin{description}
\item[$1. \Rightarrow 2.$] By Proposition \ref{prop:ercsimplieslc}, given $x \in \mathbf{X}$ we can compute some $V \in \mathcal{O}(\mathbf{X})$ and $B \in \mathcal{K}(\mathbf{X})$ with $x \in V \subseteq B$. In a computable metric space, given $x \in \mathbf{X}$ and $V \in \mathcal{O}(\mathbf{X})$, we can compute some $n \in \mathbb{N}$ such that $B(x,2^{-n}) \subseteq V$. Since computable metric spaces are Hausdorff, compact sets are closed, and thus $B(x,2^{-n}) \subseteq B$ implies $\overline{B}(x,2^{-n}) \subseteq B$. Moreover, we can compute $\overline{B}(x,2^{-n}) \in \mathcal{A}(\mathbf{X})$, and $\mathalpha{\cap} : \mathcal{A}(\mathbf{X}) \times \mathcal{K}(\mathbf{X}) \to \mathcal{K}(\mathbf{X})$ is computable for arbitrary represented spaces $\mathbf{X}$. Thus, we can obtain $\overline{B}(x,2^{-n}) = \overline{B}(x,2^{-n}) \cap B \in \mathcal{K}(\mathbf{X})$.

\item[$2. \Rightarrow 1.$] A computable metric space has a dense sequence $(x_n)_{n \in \mathbb{N}}$. We apply $\operatorname{CompactBall}$ to each to obtain $(k_n)_{n \in \mathbb{N}}$ and the sequence $(\overline{B}(x_n,2^{-k_n}) \in \mathcal{K}(\mathbf{X}))_{n \in \mathbb{N}}$. Since we can compute any $\overline{B}(x_n,2^{-i}) \in \mathcal{A}(\mathbf{X})$, and $\mathalpha{\cap} : \mathcal{A}(\mathbf{X}) \times \mathcal{K}(\mathbf{X}) \to \mathcal{K}(\mathbf{X})$ is computable; we obtain the computable double-sequence $(\overline{B}(x_n,2^{-\max \{k_n,j\}}) \in \mathcal{K}(\mathbf{X}))_{n,j \in \mathbb{N}}$. As basic open sets we use $(B(x_n,2^{-i}) \in \mathcal{O}(\mathbf{X}))_{n,i \in \mathbf{N}}$, and we set $B(x_n,2^{-i}) \ll \overline{B}(x_m,2^{-\max \{k_m,j\}})$ iff $d(x_n,x_m) + 2^{-\max \{k_m,j\}} < 2^{-i}$. It is a straight-forward calculation to verify that this fulfills the main property of an ercs.
\end{description}
\end{proof}
\end{proposition}

\begin{corollary}
\label{corr:computablycompact}
Every computably compact computable metric space admits an ercs.
\end{corollary}

\subsection{Local compactness vs $\sigma$-compactness}
A notion somewhat related to local compactness is $\sigma$-compactness. We briefly explore their relationship in the effective setting.
\begin{definition}
Call $\mathbf{X}$ effectively $\sigma$-compact, if there is a computable $(K_n \in \mathcal{K}(\mathbf{X}))_{n \in \mathbb{N}}$ with $\mathbf{X} = \bigcup_{n \in \mathbb{N}} K_n$.
\end{definition}

\begin{observation}
Every space admitting an ercs is effectively $\sigma$-compact, with the sets $(B_n \in \mathcal{K}(\mathbf{X}))_{n \in \mathbb{N}}$ of the ercs serving as witness.
\end{observation}

Being locally compact does not imply being $\sigma$-compact, as is well-known even for metric spaces. A potential counter-example works as follows:

\begin{example}
\label{ex:basicstar}
Let $\mathbf{S}$ have the underlying set $\{\infty\} \cup \{(n,x) \mid n \in \mathbb{N} \wedge x \in \uint\}$ and the metric $d$ satisfying that $d(\infty,(n,x)) = 2^{-n} + x$, $d((n,x),(n,y)) = |x - y|$ and $d((n,x),(m,y)) = x + y + 2^{-n} + 2^{-m}$ for $n \neq m$. This is readily seen to be a computable metric space, and the sets $K_\infty = \{\infty\}$, $K_n = \{n\} \times \uint$ form an effective partition into compact sets. Yet any closed ball $\overline{B}(\infty, 2^{-j})$ contains the sequence $((\ell, 2^{-j-1})_{\ell > j})$ which has no accumulation point. Hence $\overline{B}(\infty, 2^{-j})$ is never compact, and $\mathbf{S}$ not locally compact.
\end{example}

We can modify this example to yield further separating constructions. In particular, we have:

\begin{example}
There exists a computable metric space which is effectively $\sigma$-compact and locally compact, but not effectively locally compact.
\begin{proof}
Pick some effective enumeration $(\Phi_s)_{s \in \mathbb{N}}$ of \emph{some} Turing machines. For each $s$, we consider the subspace $\mathbf{S}_s$ of $\mathbf{S}$ defined by $\infty \in \mathbf{S}$ and $(n,x) \in \mathbf{S}_s$ iff $x \geq 2^{-j}$ and $\Phi_s$ writes exactly $j$ symbols when run for $n$ steps. We now consider the disjoint union $\biguplus_{s \in \mathbb{S}} \mathbf{S}_s$, which is an effectively $\sigma$-compact computable metric space again.

If the $\Phi_s$ are such that any $\Phi_s$ will write only finitely many symbols at all, then $\biguplus_{s \in \mathbb{S}} \mathbf{S}_s$ is locally compact, since it is a discrete union of countably many singletons and copies of the unit interval. For it to be effectively locally compact, we need to be able to compute the map $\operatorname{CompactBall}$ by Proposition \ref{prop:inmetricspaces}, which in particular would mean that given $s \in \mathbb{N}$ we can compute some $d \in \mathbb{N}$ such that $\overline{B}(\infty_s,2^{-d})$ is compact, where $\infty_s$ is the $s$-th copy of $\infty$. The reasoning in Example \ref{ex:basicstar} shows that that means that $\Phi_s$ will never write more than $d$ symbols. We can easily chose a family $(\Phi_s)_{s \in \mathbb{N}}$ such that each $\Phi_s$ writes only finitely many times, but such that we cannot compute an upper bound on how often from $s$.
\end{proof}
\end{example}

\subsection{Comparison to notions in the literature}
\label{subsec:comparison}
\cite[Definition 3]{zhenga} defines a computable metric space $\mathbf{X}$ to be effectively locally compact, if there is a computable positive function $\gamma : \mathbf{X} \to \mathbb{R}$ such that each $\overline{B}(x,\gamma(x))$ is compact. Proposition \ref{prop:inmetricspaces} shows that our definition implies theirs. Any compact metric space trivially satisfies their condition, but by combination of Proposition \ref{prop:compactcomputablycompact} and Corollary \ref{corr:computablycompact} a compact computable metric space admits an ercs iff it is computably compact. Thus, a compact but not computably compact computable metric space separates the two notions.

In \cite{zhengb} (the journal version of the conference paper \cite{zhenga}), the requirement is that given $x \in \mathbf{X}$ and $0 < \delta \in \mathbb{R}$, one can compute some $\overline{B}(x,\rho) \in (\mathcal{K}\wedge\mathcal{V})(\mathbf{X})$ with $\overline{B}(x,\rho) \subseteq B(x,\delta)$. We observe the following:

\begin{proposition}
The map $\operatorname{Radius}: \subseteq \mathbf{X} \times (\mathcal{K}\wedge\mathcal{V})(\mathbf{X}) \mto \mathbb{R}$ mapping $x, K$ to some $\rho \in \mathbb{R}$ such that $K = \overline{B}(x,\rho)$ is computable.
\begin{proof}
We can compute $\rho_- \in \mathbb{R}_>$ such that $\rho_- < r$ iff $K \subseteq B(x,r)$ since we have $K$ as a compact set. We can also compute $\rho_+ \in \mathbb{R}_<$ such that $r < \rho_+$ iff $\overline{B}(x,r)^C \cap K \neq \emptyset$ since we have $K$ as an overt set. Now any $\rho$ with $\rho_- \leq \rho \leq \rho_+$ is a valid answer. We can start with the assumption that $\rho_- = \rho_+$, and start computing $\mathbb{R} \ni \rho = \rho_- = \rho_+$ from that. If $\rho_- < \rho_+$, we will eventually notice, and can then output some rational in that interval.
\end{proof}
\end{proposition}

Since the $2. \Rightarrow 1.$-direction of Proposition \ref{prop:inmetricspaces} does not make use of the radius having the special form $2^{-n}$, we see that the definition of effective local compactness from \cite{zhengb} implies the existence of an ercs. The difference lies in \cite{zhengb} requiring that the closed balls are produced as compact and overt sets. In general, we can compute $\mathrm{cl} B(x,r)$ as an overt set given $x \in \mathbf{X}$ and $r \in \mathbb{R}$, but not $\overline{B}(x,r)$. Of course, for the cases that $\mathrm{cl} B(x,r) = \overline{B}(x,r)$, we can obtained the ball as both closed (respectively compact) and overt set. This is, in a way, the typical case:

\begin{lemma}[Banakh \cite{banakh}]
\label{lemma:banakh}
Let $(X,d)$ be a separable metric space and $x \in X$. Then $\{r \in \mathbb{R} \mid \mathrm{cl} B(x,r) \neq \overline{B}(x,r)\}$ is countable.
\begin{proof}
Fix a countable basis $(U_n)_{n \in \mathbb{N}}$. Let $R = \{r \in \mathbb{R} \mid \mathrm{cl} B(x,r) \neq \overline{B}(x,r)\}$. For each $r \in R$, pick some $y(r) \in \overline{B}(x,r) \setminus \mathrm{cl} B(x,r)$. Then there must be some $n(r) \in \mathbb{N}$ with $y(r) \in U_{n(r)}$ and $U_{n(r)} \cap B(x,r) = \emptyset$. If $|R| > |\mathbb{N}|$, there must be $r_1, r_2 \in R$ with $r_1 < r_2$ but $n(r_1) = n(r_2)$. Now we have that $y(r_1) \in U_{n(r_1)}$, $d(x,y(r_1)) = r_1$ and $U_{n(r_1)} \cap B(x,r_2) = \emptyset$, contradiction. So $R$ is countable.
\end{proof}
\end{lemma}

\begin{proposition}
\label{prop:smallovertball}
Let $\mathbf{X}$ be computable Polish space. The map $\operatorname{NiceRadius} :\subseteq \mathbf{X} \times \mathbb{R}^{>0} \times \mathcal{K}(\mathbf{X}) \mto \mathbb{R}$ with $(x,r,K) \in \dom(\operatorname{NiceRadius})$ iff $K = \overline{B}(x,r)$ and $r' \in \operatorname{NiceRadius}$ iff $0 < r' < r$ and $\mathrm{cl} B(x,r') = \overline{B}(x,r')$ is computable.
\begin{proof}
For each $d \in \mathbb{R}$ with $0 < d < r$ we can compute $\{y \in \mathbf{X} \mid d(x,y) = d\} \in \mathcal{K}(\mathbf{X})$, as this is trivially computable as a closed set, and we then take the intersection with the provided compact set $\overline{B}(x,r)$. Now given $n \in \mathbb{N}$ and $\{y \in \mathbf{X} \mid d(x,y) = d\} \in \mathcal{K}(\mathbf{X})$, we can semidecide if $\forall y \in \mathbf{X} \left (d(x,y) = d \Rightarrow B(x,d) \cap B(y,2^{-n}) \neq \emptyset \right )$, since this is a universal quantification over a compact set and an open predicate. From this, we see that for each $n \in \mathbb{N}$, we can obtain the open set:
$$U_n = \{d \in \mathbb{R} \mid 0 < d < r \wedge \forall y \in \mathbf{X} \left (d(x,y) = d \Rightarrow B(x,d) \cap B(y,2^{-n}) \neq \emptyset \right )\}$$
Note that $\mathrm{cl} B(x,d) = \overline{B}(x,d) \Leftrightarrow \forall n \in \mathbb{N} \ d \in U_n$. The set $D := \bigcap_{n \in \mathbb{N}} U_n$ is available to us as a $\Pi^0_2$-set, and by Lemma \ref{lemma:banakh}, it is co-countable and hence co-meager. We can thus apply the Computable Baire Category theorem \cite{brattka7} to compute some $r' \in D$.
\end{proof}
\end{proposition}

\begin{corollary}
A computable Polish space is effectively locally compact in the sense of \cite{zhengb} iff it admits an ercs.
\end{corollary}

It is not clear whether the requirements of having a surrounding compact ball, or of the space being Polish, are actually needed. We thus raise the following question:

\begin{question}
Is the map $\operatorname{OvertBall} : \mathbf{X} \times \mathbb{R}^{>0} \mto \mathcal{V}(\mathbf{X}) \times \mathbb{R}^{>0}$ defined by $(K,r') \in \name{OvertBall}(x,r)$ iff $r' < r$ and $K = \overline{B}(x,r')$, computable for all computable metric spaces? Or at least for all computable Polish spaces?
\end{question}

\cite{edalat} is considering countably-based Hausdorff spaces, and is defining effective local compactness in terms of an enumeration of a basis $(O_n)_{n \in \mathbb{N}}$ making $O_n \subseteq O_m$ and $\mathrm{cl} O_n \subseteq O_m$ decidable in $n$ and $m$; and moreover there is a cover $\mathbf{X} = \bigcup_{i \in \mathbb{N}} \mathbf{X}_i$ by compact subspaces, such that $\mathbf{X}_i \setminus O_n \subseteq O_{m_1} \cup \ldots \cup O_{m_k}$ is decidable in the indices. It is immediate that this requirement implies ours, but asking for actual containment to be decidable rather than employing formal containment makes the framework of \cite{edalat} much more restrictive.

The definition of effectively local compactness for computable Polish spaces in \cite{kamo} is asking for a cover $\mathbf{X} = \bigcup_{i \in \mathbb{N}} \mathbf{X}_i$, where $(\mathbf{X}_i \in (\mathcal{V}\wedge\mathcal{K})(\mathbf{X}))_{i \in \mathbb{N}}$ is a computable sequence, and from $x \in \mathbf{X}$ we can compute some $0 < \delta \in \mathbb{R}$ and $i \in \mathbb{N}$ such that $B(x,\delta) \subseteq \mathbf{X}_i$. From Proposition \ref{prop:inmetricspaces} we conclude that their definition implies ours. What is missing for the converse is that we only get a cover by compact sets, not by compact and overt sets. We can use Proposition \ref{prop:smallovertball} to circumvene this, and conclude that for computable Polish spaces the definitions of effective local compactness from \cite{zhengb}, from \cite{kamo} and from this paper all agree.

\section{The hyperspace $(\mathcal{A}\wedge\mathcal{V})(\mathbf{X})$}
\label{sec:av}

We move on to our application of the machinery of effective local compactness. We study the hyperspace $(\mathcal{A}\wedge\mathcal{V})(\mathbf{X})$ of sets given as both closed and overt. In the language of Weihrauch \cite{weihrauchd}, this is the full information representation of the closed sets. In constructive mathematics, the computable elements of $(\mathcal{A}\wedge\mathcal{V})(\mathbf{X})$ are often called \emph{located}.

Our main result is that whenever $\mathbf{X}$ admits an ercs, then $(\mathcal{A}\wedge\mathcal{V})(\mathbf{X})$ is computably compact and computably metrizable. This generalizes a result from  \cite{ziegler9} for computably compact computable metric spaces.

\subsection{Compactness of $(\mathcal{A}\wedge\mathcal{V})(\mathbf{X})$}
We fix a space $\mathbf{X}$ with an ercs $((U_n)_{n \in \mathbb{N}},(B_n)_{n \in \mathbb{N}},R)$.

\begin{definition}
Call $p \in \Cantor$ \emph{consistent}, if whenever $p(n) = 1$, $U_n \ll B_k$ and $B_k \subseteq U_{n_0} \cup \ldots \cup U_{n_\ell}$, then $p(n_i) = 1$ for some $i \leq \ell$.
\end{definition}

\begin{theorem}
\hfill
\begin{enumerate}
\item The set of consistent $p$ is $\Pi^0_1$.
\item If for some set $A$ we have that $p(n) = 1 \Leftrightarrow A \cap U_n \neq \emptyset$, then $p$ is consistent.
\item If $p$ is consistent, then $\{U \in \mathcal{O}(\mathbf{X}) \mid \exists n, \ell \in \mathbb{N} \ p(n) = 1 \wedge U_n \ll B_\ell \wedge B_\ell \subseteq U\}$ uniformly defines an overt set $A \in \mathcal{V}(\mathbf{X})$.
\item Moreover, we have that $\mathbf{X} \setminus A = \bigcup_{\{n \mid p(n) = 0\}} U_n$.
\end{enumerate}
\begin{proof}
\hfill \begin{enumerate}
\item We can enumerate all conditions that need to be fulfilled, and decide for any enumerated condition whether it is indeed fulfilled.
\item If $p(n) = 1$, then there exists some $x \in A \cap U_n$. If then $U_n \ll B_k$ and $B_k \subseteq U_{n_0} \cup \ldots \cup U_{n_\ell}$, then in particular $U_n \subseteq U_{n_0} \cup \ldots \cup U_{n_\ell}$, hence there is some $i \leq \ell$ with $x \in U_{n_i}$. By assumption, it follows that $p(n_i) = 1$, i.e.~$p$ is consistent.
\item Given some consistent $p$, consider the set $A = \{x \in \mathbf{X} \mid \forall n \in \mathbb{N} \ x \in U_n \rightarrow p(n) = 1\}$. We claim that for any open set $U$, we have that $\exists n, \ell \in \mathbb{N} \ p(n) = 1 \wedge U_n \ll B_\ell \wedge B_\ell \subseteq U$ iff $U \cap A \neq \emptyset$.

    Assume that $x \in U \cap A$. With the argument in Proposition \ref{prop:ercsimplieslc}, given $x \in U$ we can find $n, \ell \in \mathbb{N}$ such that $x \in U_n$, $U_n \ll B_\ell$ and $B_\ell \subseteq U$. Since $x \in A$, from $x \in U_n$ we can conclude $p(n) = 1$, and conclude this direction of the argument.

    For the other direction, we assume that $\exists n, \ell \in \mathbb{N} \ p(n) = 1 \wedge U_n \ll B_\ell \wedge B_\ell \subseteq U$ yet $U \cap A = \emptyset$, and seek to derive a contradiction. It immediately follows that $B_\ell \cap A = \emptyset$. By definition of $A$, that means that for each $x \in B_\ell$ there exists some $U_{n(x)}$ with $x \in U_{n(x)}$ and $p(n(x)) = 0$. Since $B_\ell \subseteq \bigcup_{x \in B_\ell} U_{n(x)}$ and $B_\ell$ is compact, there are finitely many $n_0,\ldots,n_j$ chosen amongst the $n(x)$ such that $B_\ell \subseteq U_{n_0} \cup \ldots \cup U_{n_j}$. Together with $U_n \ll B_\ell$, $p(n) = 1$ and consistency of $p$ we would get that $p(n_i) = 1$ for some $i \leq j$, contradicting the choice of the $n_i$.
\item If $x \in U_n$ and $p(n) = 0$, then it follows by construction that $x \notin A$. Conversely, if $x \notin A$, then there must be some $n \in \mathbb{N}$ with $x \in U_n$ and $p(n) = 0$.
\end{enumerate}
\end{proof}
\end{theorem}

\begin{corollary}
If $\mathbf{X}$ admits an ercs, then there is a $\Pi^0_1$-subset $A$ of $\Cantor$ such that $\mathcal{V}(\mathbf{X})$ is computably isomorphic to $\mathrm{EC}^{-1}(A) \subseteq \mathcal{O}(\mathbb{N})$, where $\mathrm{EC} : \mathcal{O}(\mathbb{N}) \to \Cantor$ maps an enumeration of a set to its characteristic function.
\end{corollary}

\begin{corollary}
If $\mathbf{X}$ admits an ercs, then there is a $\Pi^0_1$-subset $A$ of $\Cantor$ and a computable surjection $\psi : A \to (\mathcal{A} \wedge \mathcal{V})(\mathbf{X})$.
\end{corollary}

\begin{corollary}
\label{corr:ercscompact}
If $\mathbf{X}$ admits an ercs, then $(\mathcal{A} \wedge \mathcal{V})(\mathbf{X})$ is computably compact.
\end{corollary}

\subsection{Metrizability $(\mathcal{A}\wedge \mathcal{V})(\mathbf{X})$}

\begin{proposition}
$\mathalpha{\not\subseteq} : \mathcal{V}(\mathbf{X}) \times \mathcal{A}(\mathbf{X}) \to \mathbb{S}$ is computable.
\begin{proof}
Note that $A \not\subseteq B$ is equivalent to $A \cap B^C \neq \emptyset$, and the latter is semidecidable by definition of $\mathcal{V}$ and $\mathcal{A}$ in terms of $\mathcal{O}$.
\end{proof}
\end{proposition}

\begin{corollary}
\label{corr:avhausodrff}
For arbitrary $\mathbf{X}$, $(\mathcal{A}\wedge\mathcal{V})(\mathbf{X})$ is computably Hausdorff.
\end{corollary}

\begin{proposition}
\label{prop:tomegaembedding}
Let $\mathbf{X}$ admit an ercs. Then $(\mathcal{A}\wedge\mathcal{V})(\mathbf{X})$ embeds into $\mathbb{T}^\omega$.
\begin{proof}
Fix an $((U_n)_{n \in \mathbb{N}},(B_n)_{n \in \mathbb{N}},R)$. We define the computable embedding $\phi : (\mathcal{A}\wedge\mathcal{V})(\mathbf{X}) \to \mathbb{T}^\omega$ as follows: If $U_n \cap A \neq \emptyset$, then $\phi(A)(n) = 1$ -- we can recognize this having access to $A \in \mathcal{V}(\mathbf{X})$. If there exists some $\ell \in \mathbb{N}$ with $B_\ell \cap A = \emptyset$ (which we can recognize by knowing $A \in \mathcal{A}(\mathbf{X})$) and $U_n \ll B_\ell$, then $\phi(A)(n) = 0$. Else $\phi(A)(n) = \bot$.

To compute the inverse of $\phi$, first note that we can recover $A \in \mathcal{A}(\mathbf{X})$ by noting that $A^C = \bigcup_{\{n \mid \phi(A)(n) = 0\}} U_n$. Then, note that by the main property of an ercs, we have that $U \cap A \neq \emptyset$ for open $U$ iff $\exists n, \ell \in \mathbb{N}$ such that $U \supseteq B_\ell$, $U_n \ll B_\ell$ and $U_n \cap A \neq \emptyset$. Using this twice, we can recover $A \in \mathcal{V}(\mathbf{X})$ from $\phi(A) \in \mathbb{T}^\omega$ via $U \cap A \neq \emptyset$ iff $\exists n, \ell \in \mathbb{N} \ \phi(A)(n) = 1 \wedge U_n \ll B_\ell \wedge B_\ell \subseteq U$.
\end{proof}
\end{proposition}

\begin{corollary}
\label{corr:ercscountablybased}
Let $\mathbf{X}$ admit an ercs. Then $(\mathcal{A}\wedge\mathcal{V})(\mathbf{X})$ admits an effective countable basis.
\end{corollary}

By combining Corollaries \ref{corr:ercscompact}, \ref{corr:avhausodrff} and \ref{corr:ercscountablybased}, we see that whenever $\mathbf{X}$ admits an ercs, then $(\mathcal{A}\wedge\mathcal{V})(\mathbf{X})$ is a computably compact computably Hausdorff effectively countably-based space. It was shown in \cite{paulytsuiki-arxiv} (using Schr\"oder's effective metrization theorem \cite{schroder8,grubba3}) that these conditions together imply computable metrizability. We thus get:

\begin{corollary}
\label{corr:metrizable}
Let $\mathbf{X}$ admit an ercs. Then $(\mathcal{A}\wedge\mathcal{V})(\mathbf{X})$ is computably metrizable.
\end{corollary}

\hide{
\subsection{Overtness of $(\mathcal{A}\wedge\mathcal{V})(\mathbf{X})$}

\begin{proposition}
Let $\mathbf{X}$ admit an ercs, be computably Hausdorff and computably overt. Then $(\mathcal{A}\wedge\mathcal{V})(\mathbf{X})$ is computably overt.
\begin{proof}
Consider the inverse of the embedding of $(\mathcal{A}\wedge\mathcal{V})(\mathbf{X})$ into $\mathbb{T}^\omega$ provided in Proposition \ref{prop:tomegaembedding}. To show that $(\mathcal{A}\wedge\mathcal{V})(\mathbf{X})$ is computably overt, it suffices to show that we can enumerate the basic open sets in $\mathbb{T}^\omega$ intersection the domain of that inverse.

Any basic open set in $\mathbb{T}^\omega$ indicates requirements of the form that $B_\ell \cap A \neq \emptyset$ whenever $U_i \ll B_\ell$ for $i \in I$, $I$ finite; and that $U_j \cap A = \emptyset$ for $j \in J$, $J$ finite. THIS DOES NOT WORK LIKE THIS
\end{proof}
\end{proposition}}

\subsection{Related work in classical topology}
\cite{hola} shows that for a Hausdorff space $\mathbf{X}$, the Fell topology is normal iff $\mathbf{X}$ is Lindel\"of and locally compact; and \cite{beer2} shows that for Hausdorff $\mathbf{X}$, the Fell topology is Hausdorff iff $\mathbf{X}$ is locally compact. Compactness of the Fell topology, however, holds for arbitrary Hausdorff spaces \cite{beer}. We thus see the precise opposite for the behaviour of the Fell topology and of the space $(\mathcal{A}\wedge\mathcal{V})(\mathbf{X})$. This perplexing feature shall serve as a reminder that the hyperspace constructions in our work are exploiting the cartesian-closure of our ambient category, and thus do not apply to $\mathcal{TOP}$.

\section*{Acknowledgements}
This research was partially supported by the Royal Society International Exchange Grant 170051 ``Continuous Team Semantics: On dependence and independence in a continuous world''.

The author is grateful to Matthew de Brecht for enlightening discussion.

\bibliographystyle{eptcs}
\bibliography{../../spieltheorie}

\end{document}